\documentclass[preprint,showpacs,prl,aps]{revtex4-1}
\usepackage{graphicx}
\usepackage{epsfig}
\usepackage{color}
\usepackage{cancel}
\usepackage{dcolumn}
\usepackage{bm}

\begin{document}
\title{Demystifying Quantum Mechanics}

\author{Ana Elisa D. Barioni}
\author{Felipe B. Mazzi}
\author{Elsa Bifano Pimenta}
\author{Willian Vieira dos Santos}
\author{Marco A. P. Lima}
\affiliation{Instituto de F\'{\i}sica ``Gleb Wataghin'', Universidade Estadual de
Campinas, 13083-859, Campinas, SP, Brazil}

\date{\today}

\begin{abstract}
Why does such a successful theory like Quantum Mechanics have so many mysteries? The history of this theory is replete with dubious interpretations and controversies, and yet a knowledge of its predictions, however, contributed to the amazing technological revolution of the last hundred years.  In its very beginning Einstein pointed out that there was something missing, due to contradictions with the relativity theory. So, even though Quantum Mechanics explains all the nanoscale physical phenomena, there were many attempts to find a way to ``complete" it, e.g. hidden-variable theories. In this paper, we discuss some of those enigmas, with special attention to the concepts of physical reality imposed by quantum mechanics, the role of the observer, prediction limits, a definition of collapse, and how to deal with correlated states (the basic strategy for quantum computers and quantum teleportation). That discussion is carried out within the framework of accepting that there is in fact nothing important missing, rather we are just restricted  by the limitations imposed by quantum mechanics. The mysteries are thus explained by a proper interpretation of those limitations, which is achieved by introducing  two interpretation rules within the Copenhagen paradigm.
\end{abstract}


\maketitle

In the series ``Cosmos: Possible Worlds"\cite{cpw},  Neil deGrasse Tyson, presenting the double slit experiment for photons, says: ``On the smallest possible scale that we've ever discovered, the quantum universe, the mere act of observation changes reality". This interpretation can be misleading, but it is very present in physics classes all over the world. Indeed, it brings us back to the well considered and famous Schr{\"o}dinger's cat dilemma: can the cat be in a mixed state of dead and alive? The interpretation given by the Cosmos show, seen by millions of people, by itself justifies the discussion below. Furthermore, one of the most puzzling mysteries of quantum mechanics is in entangled particle systems. For these systems,   Einstein created the term ``spooky action at a distance", which can be put in the following terms, if information is instantly transmitted between correlated particles, it would be in direct confrontation with special relativity theory. This was published in Einstein et al.'s famous article~\cite{EPR}, entitled ``Can Quantum-Mechanical Description of Physical Reality Be Considered Complete?", followed by one with the same title by N. Bohr~\cite{NB}, defending a different vision.  Another motivation for this paper is examining the mystery of collapses of wave functions in quantum mechanics. \par

All the aforementioned difficulties in interpreting the outcomes of quantum mechanics inspired several attempts to complete the theory. We mention here only some of them, such as another famous paper by D. Bohm, entitled ``A Suggested Interpretation of the Quantum Theory in Terms of ‘Hidden’ Variables”~\cite{BI},   the Many-Words interpretation~\cite{MWI}, and more recently, the QBism interpretation~\cite{QBI1,QBI2}.  In 1964, for entangled particles, Bell~\cite{BT} proposed an inequality theorem, based on the hidden variables assumption in a paper entitled ``On the Einstein-Podolsky-Rosen Paradox", and later experiments~\cite{Bex} have shown that Quantum Mechanics in fact gives the right predictions, but this did not stop the interpretation discussions. A nice debate about these interpretations (just to mention that the mysteries interpretations are still very much ``alive")  can be found in the World Science Festival, ``Measure for Measure: Quantum Physics and Reality", a debate moderated by Brian Greene with David Z. Albert, Sean Carroll, Sheldon Goldstein, and Ru\"ediger Schack~\cite{WSF}. 
 \par 

Quantum Mechanics is a very well established theory, and we will assume that the reader is familiar with  some of the most common text books that describe it. References \cite{book1}-\cite{book5} are a few examples (used  at our university), but there exist many others. All these books follow what is known as the Copenhagen paradigm. We will not make a revision of the theory here, and  we will cite and use known aspects of it without any formality. \par

The historical heart of the theory can be represented by the time-dependent Schr{\"o}dinger equation  for a particle:

\begin{equation}
i\hbar \frac{\partial \Psi ({\vec r}, t)}{\partial t}= H\Psi({\vec r}, t)\; ,
\label{SE}
\end{equation}
where $\psi ({\vec r}, t)$ is the wave function of the particle  in the time $t$ and $H$ is the Hamiltonian (it describes the environment of the particle). The wave function is a map of possibilities of what can happen to the particle.  This equation allows the prediction of the future of the wave function, $\psi ({\vec r}, t)$, at $t$, if we know this wave function, $\psi ({\vec r}, t_0)$, at a particular time $t_0$.  The idea of predicting the future of a particle, knowing its environment and its initial state  (position and velocity of the particle at time $t_0$), came from Isaac Newton's classical equations (which can be  written in different forms, as the Lagrangian equations, derived from the principle of least action, the Hamilton-Jacobi equations, and others). Wave equations, like Eq.~\ref{SE}, spoil the classical logic. Uncertainty relations, inherent to wave equations,  prohibit the precise knowledge of the position and velocity of a particle at the same time.  The Schr{\"o}dinger equation predicts no specific trajectories, only combinations of all possible trajectories (like in the Feynman formalism~\cite{book1}). The most you can get is  the knowledge of the wave function in the future and everything you can extract from it. To solve Eq.~\ref{SE} you need the Hamiltonian and the wave function in a particular time of your choice (the initial state). Academic Hamiltonians are relatively easy to assemble and the task is made even easier if the quantum problem has a classical analog. Real Hamiltonians, containing all sorts of traps and structures of the real world that can interact with the particle, are often too difficult to write so that we impose simplifications taking us towards the academic Hamiltonians, but always aiming to be good approximations of the real Hamiltonians. On the other hand, the initial state  is impossible to determine without you or somebody (or something) interfering in the system. The collapse of the wave function, as being due to measurement, is well accepted in the Copenhagen paradigm of Quantum Mechanics.  \par 

This leads us to our first  interpretation rule: there is no way you can determine the initial state of a particle beforehand, unless (i)  you have interfered in the system,  by making a measurement of a particular observable that causes the collapse of the original state to an eigenstate of your observable or (ii) because nature (the environment)  has collapsed~\cite{NC}  the original state to a particular  eigenstate of some observable, and you know which one it is. Either way you have no idea about the original state (the state prior to any of these collapses) of the system.  \par 

A good example for this rule is the oven of silver atoms in the Stern-Gerlach experiment~\cite{SG}. There is nothing you can do to learn which spin state a given silver atom is in, when it leaves the oven. Any attempt to measure the spin of the atom will collapse the original state into an eigenstate associated with the value of your measurement. The only information acquired about the original state, after you have performed a measurement, is that it had the component associated with your result. As discussed further on, this first rule will explain the spooky action at a distance mystery~\cite{EPR} of  correlated (entangled) particles. \par

You could argue that you know how to get the stationary states from Eq.~\ref{SE}, and that the future of these states is very simple to predict since  $\psi ({\vec r}, t)=e^{-\frac{i E}{\hbar}(t-t_0)}\psi ({\vec r}, t_0)$. Indeed, systems in stationary states are in a kind of quantum mechanics frozen reality. This follows as the wave functions at different times differ among themselves only by a phase factor, that does not affect any prediction about the system. Hence, if you know that the system has collapsed to a particular stationary state, it will be there forever (unless the environment acts on it, like the broadening of the electronic lines of an excited atom, due to local and non-local effects,  that will eventually take the system to its ground state - in this case you have a collapse due to the environment that nobody needs to observe for it to happen). Stationary states, even for continuum states, might thus be considered as boring because they represent a frozen situation. Consider, for example, the free-particle wave function $\psi ({\vec r}, t)=e^{-\frac{i E}{\hbar}(t-t_0)}e^{i {\vec p}\cdot {\vec r}/\hbar}$. This free-particle solution is obtained by assuming a very simplified scenario, in which there is nothing in the universe besides the free particle and that the space is homogeneous. Even the probability flux of this situation behaves as a homogeneous river pointing in the ${\vec k}$ direction. The situation gets more elaborate if you look for a solution for the scattering of this particle against a target with structure (i.e. possessing internal degrees of freedom). In this case, you can get probability fluxes for all processes due to the collision of the particle against the structured target. For this you have to include  the structured target and its interaction with the particle in the Hamiltonian, with the solution showing that the outgoing particle may leave the target in any of the energetically possible states. You don't know which process will take place, but you know the probabilities (related to the cross sections) of each one. \par 

If you want to develop intuition for more abrupt changes in reality (with respect to time), such as collapses of the wave function, you must first place the particle in a particular region and throw it against the structured target, for instance, the apparatus of a two slit interference experiment. For this you need a combination of free solutions (which is also a solution), where the combination places the particle within a wave packet centered at a particular position. This packet center, not the particle, follows a classical trajectory (the same as the classical free particle), and the particle, of course, can only be found inside the wave packet. For a proper description of this problem, the Hamiltonian of the system must contain the barrier and the free passage through the two slits. Suppose we have sufficient detectors (described in our Hamiltonian),  so that the particle can be detected if it bounces back or if it hits any place of the barrier. This set of detectors is thus triggered only if the particle does not go through one of the slits. If the particle's collision against the detectors causes a click, the absence of clicks would mean that the incident wave packet became two small packets emerging from the slits. If you now place a sensitive screen in front of these combined wave packets, you will produce a mark in this screen. Where? Wherever the combination of the two packets is different from zero. For each particle, we will have only one mark on the screen. The repetition of this experiment, however, ultimately gives the well known interference pattern of the double-slit experiment and it appears only if the wave packet goes through both slits. If the act of an observer were to block one of the slits, the interference pattern would disappear. If a particle interacts with the environment (suppose that this may happen only when it passes through one of the slits) the interference pattern also disappears, no matter if it is being observed or not. The interaction with the environment represents a collapse, which ensures that the particle really passed through one of the slits. Hence the mark on the screen of a single particle experiment represents another collapse of the wave function. If it happens in one place, it cannot happen in any other place. There are no interference effects without diffraction. So the observer in the  double-slit experiment must be at least destroying (reshaping)   the diffraction pattern, due to each slit, in order to remove the interference pattern. This can be done by focusing the beam, or something else, that changes the photon/electron  wave packet or/and the Hamiltonian. \par

This leads us to our second important interpretation rule, which is: the wave function is just a map of possibilities and probabilities. The environment (described by the Hamiltonian) shapes the wave packet, but the wave packet does not change the environment. Only the particle (inside the packet) can interact with the environment, and when that happens you have a collapse of the wave packet. The map (this human invention) of probabilities and possibilities zeroes out  everywhere, except in the region where the interaction with the environment took place.\par

Good examples~\cite{book1,G} of Hamiltonians shaping the packets (in this case split packets), are the Aharonov-Bohm effect experiment and the gravity-induced quantum interference experiment. Now, let us use the above discussion to reinterpret the Schr\"odinger's cat experiment, to discuss the phenomenon of correlation between two particles, and take a short look at the collapses. \\
{\bf Schr\"odinger's cat: } Let us look at this with the help of the Stern-Gerlach experiment. We have a poor cat inside a closed cage, tubed to receive gas from two possible triggers. Outside we have an oven expelling silver atoms and a Stern-Gerlach apparatus. One single atom enters the apparatus, and we have no a priori idea if it will move up or down along the strongly varying magnetic field. After going through the apparatus the associated wave packet is split in two. If the particle  goes up, it triggers (very sensitive multiplier) a cat enjoyable perfume. If it goes down, it conversely triggers the Schr\"odinger deadly poison gas. Our dynamic map says that the split packet hits the two triggers at about the same time, but only one of them contains the particle. Only the one containing the particle (rule 2) will trigger the respective gases, the other packet piece will disappear with the collapse. Thus the cat has never been in a mixed dead/alive state, namely it will be either dead or alive depending on the spin state of the silver atom. Therefore the wave function just represents a map of probabilities. When your detector finds a particle you don't even know if you had a split packet before the detection. The collapse can happen instantly, respecting special relativity, because it does not carry any information. Although you can induce collapses, exact predictions of where it is going to happen  are not defined by Eq.~\ref{SE}. This equation only furnishes the quantum mechanics probabilistic rules. Hence the theory seems to be complete, so that we just have to accept these imposed limitations to predict the future. \\ 

{\bf Correlation between two particles: } The simplest case is a positronium atom in its ground state (singlet state, total spin equals zero - if one particle has spin up, in any direction, the other has spin down in the same direction). If the atom is now taken apart (ionized), without changing the total angular momentum, we have the well-established situation: if the positron has spin up (down) the electron has spin down (up), as long as they are measured in the same direction.  The first important question derived from the above discussion is: after the particles have been separated, can you tell that the system is really correlated by making any kind of measurement on the positron or electron particles? According to our interpretation rule 1, the answer is NO. If you measure one particle with spin up and the other spin down, your only conclusion is that this collapsed state was part of the original state. When you measure the spin of one of the particles, the result you get is one of the results you would get from an uncorrelated system. You cannot say anything about the other particle, without being sure that the system is correlated. This argument by itself addresses Einstein's concerns about the non-locality of quantum mechanics, as no message (information) is transmitted. The preparation of the system involves a collapse with the following properties: (1) the overall singlet state ensures that one spin particle is up and the other is down, if measured in the same direction; (2) the separation ensures that only one particle will be found on each location. The map (initial state) could be represented by two wave packets of spin up and down on each location (for any, but the same, direction). Note that if the direction of measurement is not the same, any information revealed by the collapse becomes irrelevant to the other measurement. For a given direction, if you find spin up at one location, the  spin down packet disappears at that location (i.e. there can be only one particle on each side). The packet with spin up at the other location also disappears (as one is spin up the other is spin down). The new map, due to the measurement (new collapse), is the spin down on the other side and it must contain a particle. 
If you want to consider the particle separation process, imagine the pair being created in the same position (i.e. a positronium formation in the singlet state). This state~\cite{book1} can be written as:
\begin{equation}
 |{\rm spin \ singlet}\rangle=\frac{1}{\sqrt{2}}(|\hat {\bf z}+;\hat {\bf z}-\rangle-|\hat {\bf z}-;\hat {\bf z}+\rangle),
\label{singlet1}
\end{equation}
and has the same form for any orientation. Suppose now a general rotation represented by $\theta$ around the $y$-axis and $\varphi$ around the $z$-axis. This gives that the same state (differing at most by a phase factor) can be written, with the help of the kets, in the $\hat{\bf n} (\theta,\varphi)$ direction, by:
\begin{equation}
 |{\rm spin \ singlet}\rangle=\frac{1}{\sqrt{2}}(|\hat {\bf n}+;\hat {\bf n}-\rangle-|\hat {\bf n}-;\hat {\bf n}+\rangle).
\label{singlet2}
\end{equation}

This analysis shows that~\cite{UP}, at the birth of a singlet state of positronium (where the two particles are in the same position),  if you measure the spin of one particle, in any direction, the other particle will have  the opposite spin in the same direction. If you now impose a Hamiltonian that separates the particles with no spin dependencies, the spin map will be preserved.
The change in the probability map, caused by a measurement on one of the separated particles, collapses the wave function  instantaneously, without any information transmitted between the particles. Besides, you can only store and retrieve important information from this situation if you know that the system is correlated and that no dissipative Hamiltonian's act on it. This lies at the heart of quantum computing and quantum teleportation.\\

{\bf Collapses: }
Finally, let us understand that collapses happen all the time without any observer. In the beginning of the universe we had a hot plasma, and eventually electrons and protons, with their huge wave packets, found each other to form Hydrogen atoms. Nobody was observing this. If you ``throw'' a high energy electron against a gas or a surface, it will ionize along its path. Every collision can cause a collapse of the wave function until the electron is stopped and trapped, forming an atomic or molecular anion (the final situation if nothing else happens). In the case of Schr\"odinger's cat, the collapse does not occur when the box is opened, but in the fact that, invariably, the silver atom hits only one of the triggers and the state is collapsed. Gravitational fields reshape wave packets all the time (star formation is a good example), and again, no observer is necessary. The observer in the double slit experiment can destroy the interference pattern by acting on the map of probabilities and/or causing collapses, but it is important to remember that nature (the environment) does this all the time and quantum mechanics imposes tough but clear rules to predict where they can be.\\
{\bf Conclusion: }
Einstein said: ``I cannot seriously believe in (The Quantum Theory) because it cannot be reconciled with the idea that physics should represent a reality in time and in space, free from spooky actions at a distance". However, the brake of Bell's inequality theorem supported the Quantum Mechanics predictions. Reality in time and space may suggest that a possible hidden variable theory could reveal a particular path between two points, where this path is one of the (all possible) paths given by the Feynman formalism. The path integral formalism is only equivalent to the Schr{\"o}dinger equation if all possible paths (in the probability map) are taken into account. We therefore conclude that quantum mechanics is complete. Indeed we only need to accept the idea that we cannot know everything. Our interpretation rules 1 and 2 support this concept, clarifying the so-called mysteries
and thus helping to demystify quantum mechanics.\par
 
The first four authors are graduate students and their names appear in alphabetical order. F.B.M. and M.A.P.L. acknowledge support from the Brazilian agency Conselho Nacional de Desenvolvimento Cient\'\i fico e Tecnol\'ogico (CNPq) and A.E.D.B.  from the  Brazilian agency Funda\c c\~ao de Amparo \`a Pesquisa do Estado de S\~ao Paulo (FAPESP). The authors are grateful to Profs.  Marcus Aguiar (UNICAMP, ResearcherID C-9137-2012) and Michael Brunger (Flinders University, ResearcherID D-4733-2013) for  their critical reading of the manuscript and constructive comments and suggestions.



\begin{thebibliography}{99}
\bibitem{NC} In the discussion about collapses, we give several examples.
\end{thebibliography}
\end{document}